\documentclass[11pt]{article}
\usepackage{amssymb}

\usepackage[totalwidth=460truept,totalheight=600truept]{geometry}
\usepackage{latexsym,amssymb,amsmath,graphicx}
\usepackage[T1]{fontenc}

\global\arraycolsep=1truept

\begin{document}

\hfill IFUP-TH 2009/23

\vskip 1.4truecm

\begin{center}
{\huge \textbf{Renormalization Of High-Energy }} \vskip .4truecm {\huge 
\textbf{Lorentz Violating Four Fermion Models}}

\vskip 1.5truecm

\textsl{Damiano Anselmi and Emilio Ciuffoli}

\textit{Dipartimento di Fisica ``Enrico Fermi'', Universit\`{a} di Pisa, }

\textit{and INFN, Sezione di Pisa,}

\textit{Largo Pontecorvo 3, I-56127 Pisa, Italy, }

damiano.anselmi@df.unipi.it, emilio.ciuffoli@df.unipi.it

\vskip 2truecm

\textbf{Abstract}
\end{center}

{\small We study the one-loop renormalization of high-energy Lorentz
violating four fermion models. We derive general formulas and then consider
a number of specific models. We study the conditions for asymptotic freedom
and give a practical method to determine the asymptotic-freedom domain. We
also point out that in some models the RG flow contains ``rational''
Zimmermann trajectories that might hide new symmetries.}

\vskip 1truecm

\vfill\eject

\section{Introduction}

\setcounter{equation}{0}

Although Lorentz symmetry is one of the most precise symmetries in nature 
\cite{kostelecky}, the possibility that it might be violated at high
energies or very large distances is still open and has been extensively
investigated. A Lorentz symmetry violation at high energies allows us, among
the other things, to renormalize vertices that are otherwise
non-renormalizable. This result is achieved using a modified power counting
criterion, which weights space and time differently \cite{halat}. Modified
dispersion relations improve the large-momentum behavior of propagators in
such a way that, in the common perturbative framework, the theory remains
unitary, local, polynomial and causal.

Using this knowledge, it is possible to formulate a Standard Model extension 
\cite{lvsm,noh} that is CPT\ invariant, but violates Lorentz symmetry at
high energies, and contains two scalar-two fermion vertices, as well as four
fermion vertices, at the fundamental level. The inclusion of CPT violating
terms is also possible. Four fermion vertices are important for a variety of
reasons. On the one hand, they can be used to explain proton decay. On the
other hand, they can trigger a Nambu--Jona-Lasinio mechanism and give masses
to fermions and gauge fields even if the elementary Higgs boson is
suppressed \cite{noh}. In its simplest version, the scalarless Lorentz
violating Standard Model schematically reads 
\begin{equation}
\mathcal{L}_{\mathrm{noH}}=\mathcal{L}_{Q}+\mathcal{L}_{\text{kin}%
f}-\sum_{I=1}^{5}\frac{1}{\Lambda _{L}^{2}}g\bar{D}\bar{F}\,(\bar{\chi}_{I}%
\bar{\gamma}\chi _{I})+\frac{Y_{f}}{\Lambda _{L}^{2}}\bar{\chi}\chi \bar{\chi%
}\chi -\frac{g}{\Lambda _{L}^{2}}\bar{F}^{3},  \label{noH}
\end{equation}
where the quadratic terms are 
\begin{eqnarray}
\mathcal{L}_{Q} &=&\frac{1}{4}\sum_{G}\left( 2F_{\hat{\mu}\bar{\nu}}^{G}F_{%
\hat{\mu}\bar{\nu}}^{G}-F_{\bar{\mu}\bar{\nu}}^{G}\tau ^{G}(\bar{\Upsilon}%
)F_{\bar{\mu}\bar{\nu}}^{G}\right) ,  \nonumber \\
\mathcal{L}_{\text{kin}f} &=&\sum_{a,b=1}^{3}\sum_{I=1}^{5}\bar{\chi}_{I}^{a}%
\hspace{0.02in}i\left( \delta ^{ab}\hat{D}\!\!\!\!\slash -\frac{b_{0}^{Iab}}{%
\Lambda _{L}^{2}}{\bar{D}\!\!\!\!\slash}\,^{3}+b_{1}^{Iab}\bar{D}\!\!\!\!%
\slash \right) \chi _{I}^{b}  \label{lkinf}
\end{eqnarray}
and the vertices are denoted symbolically, namely without listing all
possible field differentiations and index contractions. In our notation hats
are used to denote time, bars to denote space. The weight of time is equal
to $-$1, while the weight of each space coordinate is $-$1/3. The gauge
couplings $g$ have weight 1/3. The weighted dimension of space-time is 2, so
the Lagrangian contains only terms of weights $\leqslant 2$. Moreover, $\chi
_{1}^{a}=L^{a}=(\nu _{L}^{a},\ell _{L}^{a})$, $\chi
_{2}^{a}=Q_{L}^{a}=(u_{L}^{a},d_{L}^{a})$, $\chi _{3}^{a}=\ell _{R}^{a}$, $%
\chi _{4}^{a}=u_{R}^{a}$ and $\chi _{5}^{a}=d_{R}^{a}$, $\nu ^{a}=(\nu
_{e},\nu _{\mu },\nu _{\tau })$, $\ell ^{a}=(e,\mu ,\tau )$, $u^{a}=(u,c,t)$
and $d^{a}=(d,s,b)$. The sum $\sum_{G}$ is over the gauge groups $SU(3)_{c}$%
, $SU(2)_{L}$ and $U(1)_{Y}$. Finally, $\bar{\Upsilon}\equiv -\bar{D}%
^{2}/\Lambda _{L}^{2}$, where $\Lambda _{L}$ is the scale of Lorentz
violation, and $\tau ^{G}$ are polynomials of degree 2.

The models of \cite{lvsm,noh} are anomaly-free, because gauge anomalies
cancel out exactly as in the Standard Model \cite{lvsm}. The ``boundary
conditions'' such that Lorentz invariance is recovered at low energies are
that $b_{1}^{Iab}$ tend to $\delta ^{ab}$ and $\tau ^{G}$ tend to 1 (one
such condition can be trivially fulfilled normalizing the space coordinates $%
\bar{x}$).

An important consequence of the high-energy Lorentz violation is that all
gauge interactions are super-renormalizable, therefore asymptotically free.
Moreover, since fermions have weight 1/2, the four fermion interactions are
strictly renormalizable. At energies much larger than $\Lambda _{L}$ vectors
decouple and the model (\ref{noH}) reduces to a four fermion model in two
weighted dimensions, 
\begin{equation}
\mathcal{L}_{\mathrm{HE}}=\sum_{a,b=1}^{3}\sum_{I=1}^{5}\bar{\chi}_{I}^{a}%
\hspace{0.02in}i\left( \delta ^{ab}\hat{\partial}\!\!\!\slash -\frac{%
b_{0}^{Iab}}{\Lambda _{L}^{2}}{\bar{\partial}\!\!\!\slash}\,^{3}\right) \chi
_{I}^{b}+\frac{Y_{f}}{\Lambda _{L}^{2}}\bar{\chi}\chi \bar{\chi}\chi ,
\label{he}
\end{equation}
plus free fields. The purpose of this paper is to study this type of model,
its one-loop beta functions and the conditions for asymptotic freedom. We
stress that if the high-energy model (\ref{he}) is asymptotically free, then
the full Standard Model extension (\ref{noH}) is, as well as its other
versions of ref.s \cite{lvsm,noh}.

We work out a method to determine the domain of asymptotic freedom in
quantum field theories with more couplings and apply it to some of our
models. Our approach is to study the asymptotic expansion of the running
couplings around the free fixed point. The domain of asymptotic freedom $%
\mathcal{D}_{\text{AF}}$ is determined by the arbitrary constants contained
in the expansion. The dimension of $\mathcal{D}_{\text{AF}}$ is equal the
number of positive eigenvalues (including multiplicities) of a certain
matrix $\mathcal{N}$, which depends only on the one-loop coefficients of the
beta functions.

Finally, we point out the presence of special RG trajectories that might
hide new symmetries. Indeed, if we apply Zimmermann's ``reduction of
couplings'' \cite{zimme} to our beta functions we find that some solutions
of the RG\ equations exhibit features that normally appear only in the
presence of hidden symmetries.

The paper is organized as follows. In section 2 we classify the four fermion
vertices and present the most general CPT- and rotation invariant four
fermion model. In section 3 we work out general formulas for its one-loop
beta functions. In section 4 we study some explicit examples. In section 5
we formulate our method to determine the domain of asymptotic freedom. In
section 6 we recall Zimmermann's reduction of couplings, and explain why
some of our models might possess hidden symmetries. Section 7 contains our
conclusions.

\section{Four fermion model}

\setcounter{equation}{0}

Using charge conjugation we can use only left-handed fermions, which we
collect into a vector $\ell _{j}$. We orthonormalize the kinetic terms $\bar{%
\ell}_{j}i\gamma ^{\mu }\hat{\partial}_{\mu }\ell _{k}$ and simultaneously
diagonalize the quadratic terms $\bar{\ell}_{j}i\gamma ^{\mu }{\bar{\partial}%
}_{\mu }{\bar{\partial}}^{2}\ell _{k}$ by means of a unitary transformation.

The most general CPT\ invariant, rotation invariant four fermion vertices
are of the form $L^{*}L^{*}LL$ and $LLLL$, plus its conjugate $%
L^{*}L^{*}L^{*}L^{*}$. We do not have to include vertices $L^{*}LLL$.
Indeed, in the four-component notation they can be constructed only with an
odd number of space-time indices, so they violate CPT. Precisely, a term $%
L^{*}LLL$ plus its Hermitian conjugate reads 
\begin{equation}
a(\bar{L}_{1}UL_{2})(\overline{L_{3}^{c}}VL_{4})+a^{*}(\bar{L}_{2}\gamma
^{0}U\gamma ^{0}L_{1})(\bar{L}_{4}\gamma ^{0}V\gamma ^{0}L_{3}^{c}),
\label{herme}
\end{equation}
where $U$ can be $\gamma ^{0}$ or $\gamma ^{i}$, $V$ can be $1$, $\sigma
^{0i}$ or $\sigma ^{ij}$, and $a$ is a constant. However, the combination (%
\ref{herme}) is CPT\ odd.

On the other hand, all combinations of terms $LLLL$, $L^{*}L^{*}L^{*}L^{*}$
and $L^{*}L^{*}LL$ are CPT even. Explicitly, we have the structures 
\[
(\bar{L}_{1}VL_{2}^{c})(\overline{L_{3}^{c}}VL_{4}),\qquad (\bar{L}%
_{1}UL_{2})(\bar{L}_{3}UL_{4}), 
\]
for $L^{*}L^{*}LL$, and 
\[
(\overline{L_{1}^{c}}VL_{2})(\overline{L_{3}^{c}}VL_{4}), 
\]
for $LLLL$, plus their Hermitian conjugates.

Using Fierz identities, we can show that every $LLLL$ vertex is Lorentz
invariant and of the form 
\begin{equation}
(\overline{L_{1}^{c}}L_{2})(\overline{L_{3}^{c}}L_{4}).  \label{ve1}
\end{equation}
Similarly, all four fermion vertices of type $L^{*}L^{*}LL$ have the form 
\begin{equation}
(L_{1}^{\dagger }L_{2})(L_{3}^{\dagger }L_{4}).  \label{ve2}
\end{equation}
In general, this structure is Lorentz violating, but the combination 
\begin{equation}
(L_{1}^{\dagger }L_{2})(L_{3}^{\dagger }L_{4})+(L_{1}^{\dagger
}L_{4})(L_{3}^{\dagger }L_{2})=(\overline{L_{1}}L_{3}^{c})(\overline{%
L_{2}^{c}}L_{4})=\frac{1}{2}(\bar{L}_{1}\gamma _{\mu }L_{2})(\bar{L}%
_{3}\gamma ^{\mu }L_{4}),  \label{ghino}
\end{equation}
is Lorentz invariant, which can be easily proved using the Fierz identity $%
\sigma _{\ \alpha \beta }^{i}\sigma _{\ \gamma \delta }^{i}=2\delta _{\alpha
\delta }\delta _{\gamma \beta }-\delta _{\alpha \beta }\delta _{\gamma
\delta }$.

At this point it is convenient to switch to the two-component spinor
notation. Because of (\ref{ve1}) and (\ref{ve2}), the most general four
fermion vertices are constructed with the contractions 
\begin{equation}
\ell _{i}^{\dagger }\ell _{j}\equiv \ell _{i}^{\dagger \alpha }\ell
_{j}^{\alpha },\qquad \ell _{k}^{T}\varepsilon \ell _{m}\equiv \ell
_{k}^{\alpha }\varepsilon _{\alpha \beta }\ell _{m}^{\beta },\qquad \ell
_{k}^{\dagger }\varepsilon \ell _{m}^{*}\equiv \ell _{k}^{*\alpha
}\varepsilon _{\alpha \beta }\ell _{m}^{*\beta },  \label{bili}
\end{equation}
$\alpha ,\beta ,\ldots $ being spinor indices and $^{T}$ denoting
transposition. The most general CPT- and rotation invariant interaction
Lagrangian reads 
\begin{equation}
\mathcal{L}_{4f}=\frac{1}{2\Lambda _{L}^{2}}\sum_{ijkm}\left[ (\ell
_{i}^{\dagger }\ell _{j})(\ell _{k}^{\dagger }\ell _{m})g_{ijkm}+(\ell
_{i}^{T}\varepsilon \ell _{j})(\ell _{k}^{T}\varepsilon \ell
_{m})f_{ijkm}+(\ell _{i}^{\dagger }\varepsilon \ell _{j}^{*})(\ell
_{k}^{\dagger }\varepsilon \ell _{m}^{*})f_{ijkm}^{*}\right] ,  \label{int}
\end{equation}
where the couplings $f$ and $g$ satisfy the symmetry properties 
\[
g_{ijkm}=g_{kmij}=g_{jimk}^{*},\qquad f_{ijkm}=f_{kmij}=f_{jikm}. 
\]

Because of (\ref{ghino}), the most general Lorentz invariant interaction
Lagrangian is (\ref{int}) if the couplings $g_{ijkm}$ satisfy the additional
symmetry property 
\[
g_{ijkm}=g_{imkj}. 
\]

Finally, the most general high-energy Lorentz violating four fermion model
that we are going to study reads 
\begin{equation}
\mathcal{L}=\sum_{j}\ell _{j}^{\dagger }i\left( \hat{\partial}+b_{j}\mathbf{%
\sigma }\cdot \mathbf{\bar{\partial}}\frac{{\bar{\partial}}^{2}}{\Lambda
_{L}^{2}}\right) \ell _{j}+\mathcal{L}_{4f},  \label{4f}
\end{equation}
where $\mathbf{\sigma }$ are the Pauli matrices. This model is
renormalizable. The quadratic terms $\bar{\chi}_{I}^{a}\hspace{0.02in}%
ib_{1}^{Iab}\bar{\partial}\!\!\!\slash \chi _{I}^{b}$ of (\ref{lkinf}) have
been switched off, because they are not important at high energies.
Nevertheless, the correlation functions at generic external momenta have no
infrared divergences in two weighted dimensions. Therefore, we can study the
model (\ref{4f}) in itself.

Assigning the axial charges $+1$ to $\ell _{j}$ and $-1$ to $\ell
_{i}^{\dagger }$, the $g$-terms are axially symmetric, while the $f$-terms
explicitly violate the axial symmetry. Suppressing the $f$-terms we obtain a
restricted model 
\begin{equation}
\mathcal{L}_{r}=\sum_{j}\ell _{j}^{\dagger }i\left( \hat{\partial}+b_{j}%
\mathbf{\sigma }\cdot \mathbf{\bar{\partial}}\frac{{\bar{\partial}}^{2}}{%
\Lambda _{L}^{2}}\right) \ell _{j}+\frac{1}{2\Lambda _{L}^{2}}%
\sum_{ijkm}(\ell _{i}^{\dagger }\ell _{j})(\ell _{k}^{\dagger }\ell
_{m})g_{ijkm},  \label{l4fr}
\end{equation}
that is still renormalizable. This restriction is interesting for merely
theoretical purposes. However, for phenomenological applications we include
the full set (\ref{ghino}) of interactions. Integrating the Higgs field out
in the Standard Model produces $f$-terms, which generate $g$-terms by
renormalization. Moreover, our Lorentz-violating theories can include four
fermion vertices that describe proton decay, and such terms are of both $f$-
and $g$-types \cite{strumia}.

It should be emphasized that a high-energy Lorentz violation does not imply
that the proton must decay. In this respect, we have two classes of
renormalizable models. The models of the first class are described by $B$%
-invariant Lagrangians. Then, $B$-violating vertices are not generated back
as counterterms by renormalization\footnote{%
The $B$-violations due to the $B+L$ anomaly are non-perturbative, and do not
affect the renormalization structure of the theory.}, so this choice is
consistent. The models of the second class contain $B$-violating four
fermion vertices at the classical level. Consistency with existing
experimental bounds on proton decay imply that the energy scale of Lorentz
violation $\Lambda _{L}$ must then be greater than or equal to 10$^{14\text{-%
}15}$GeV (see for example \cite{strumia}). At present, there is no reason to
expect that $\Lambda _{L}$ is much smaller than this value. However, if
neutrino masses have the Lorentz-violating origin suggested by one of us in 
\cite{lvsm}, namely they are explained by the renormalizable dimension-5
vertex 
\[
\frac{1}{\Lambda _{L}}(LH)^{2}, 
\]
where $H$ is the Higgs field, then $\Lambda _{L}$ could be around 10$^{14}$%
GeV, which is still compatible with the bound coming from proton decay%
\footnote{%
A difference of one or two orders of magnitude can always be due to the
dimensionless couplings that multiply the vertices.}.

\section{One-loop renormalization}

\setcounter{equation}{0}

In this section we study the one-loop renormalization in the most general
model (\ref{4f}). It is convenient to use the background field method.
Replace $\ell $ with $\ell +\psi $, where $\psi $ denotes the background
field. Then expand $\mathcal{L}$ in $\psi $ and keep only the quadratic part
in $\ell $. The result can be written as 
\[
\mathcal{L}_{\psi }=\frac{1}{2}\sum_{np}\left( \ell _{n}^{\dagger },\ell
_{n}^{T}\right) (Q_{n}\delta _{np}+H_{np})\left( 
\begin{tabular}{c}
$\ell _{p}$ \\ 
$\ell _{p}^{*}$%
\end{tabular}
\right) , 
\]
where $Q_{n}$ collects the kinetic terms and $H_{np}$ is a matrix quadratic
in $\psi $. In momentum space, 
\[
Q_{n}=p^{0}+b_{n}\frac{\mathbf{p}^{2}}{\Lambda _{L}^{2}}\left( 
\begin{tabular}{cc}
$\mathbf{p}\cdot \mathbf{\sigma }$ & $0$ \\ 
$0$ & $\mathbf{p}\cdot \mathbf{\sigma }^{*}$%
\end{tabular}
\right) . 
\]
Thus, the one-loop contribution to the effective action reads 
\begin{equation}
\Gamma _{\text{1}}=-\frac{i}{2}\mathrm{tr}\ln [Q_{n}\delta _{np}+H_{np}]=%
\frac{i}{4}\mathrm{tr}[Q^{-1}HQ^{-1}H]+\text{constant}+\text{finite}.
\label{g1l}
\end{equation}
Observe that the tadpole $-i\mathrm{tr}[Q^{-1}H]/2$ vanishes, since the
propagator $Q^{-1}$ is odd in $p$. For this reason there is no wave-function
renormalization at one loop.

We use a dimensional regularization where only the dimensions of space are
continued to complex values $3-\varepsilon _{2}$. There is no reason to
continue also the time dimension, since the integrals over $\hat{p}=p^{0}$
converge. The divergent parts of just two integrals are necessary to
evaluate expression (\ref{g1l}), namely 
\[
\int \frac{\mathrm{d}p_{4}\ \mathrm{d}^{3-\varepsilon _{2}}\bar{p}}{(2\pi
)^{4}}\frac{p_{4}^{2}}{D_{n}D_{p}}=\frac{\Lambda _{L}^{2}}{4\pi
^{2}\varepsilon _{2}(|b_{n}|+|b_{p}|)}+\text{finite}, 
\]
and 
\[
\int \frac{\mathrm{d}p_{4}\ \mathrm{d}^{3-\varepsilon _{2}}\bar{p}}{(2\pi
)^{4}}\frac{1}{\Lambda _{L}^{4}}\frac{(\bar{p}^{2})^{3}}{D_{n}D_{p}}=\frac{%
\Lambda _{L}^{2}}{4\pi ^{2}\varepsilon _{2}|b_{n}b_{p}|(|b_{n}|+|b_{p}|)}+%
\text{finite}, 
\]
where 
\[
D_{i}=p_{4}^{2}+b_{i}^{2}\frac{(\bar{p}^{2})^{3}}{\Lambda _{L}^{4}}+\delta
^{2} 
\]
(in Euclidean space). The fictitious mass $\delta $ is introduced to avoid
IR problems at vanishing external momenta and can be set to zero after the
evaluation. The calculations are performed using Feynman parameters to
integrate over $p_{4}$. This isolates the pole of the $\bar{p}$-integral,
therefore the divergent part.

Using these formulas we easily find 
\[
\Gamma _{\text{1}}=\frac{\Lambda _{L}^{2}}{(4\pi )^{2}\varepsilon _{2}}%
\sum_{np}t_{np}\left\{ \mathrm{tr}[H_{np}H_{pn}]-\frac{s_{np}}{3}\mathrm{tr}%
\left[ \left( 
\begin{tabular}{cc}
$\sigma ^{i}$ & $0$ \\ 
$0$ & $\sigma ^{i*}$%
\end{tabular}
\right) H_{np}\left( 
\begin{tabular}{cc}
$\sigma ^{i}$ & $0$ \\ 
$0$ & $\sigma ^{i*}$%
\end{tabular}
\right) H_{pn}\right] \right\} . 
\]

We can convert this formula from the dimensional regularization to a
conventional cut-off $\bar{\Lambda}$ on the $\bar{p}$-integral replacing $%
1/\varepsilon _{2}$ with ln$\bar{\Lambda}+$constant. Now, since $\bar{\Lambda%
}$ has weight 1/3 and, by definition, the high-energy dynamical scale $\mu $
has weight 1, matching weights and dimensions we find the identification 
\[
\frac{1}{\varepsilon _{2}}=\frac{1}{3}\ln \frac{\bar{\Lambda}^{3}}{\mu
\Lambda _{L}^{2}}. 
\]
The relation between the one-loop bare and renormalized Lagrangians is 
\[
\mathcal{L}_{\mathrm{B}}=\mathcal{L}_{\mathrm{R}}-\Gamma _{1}. 
\]
The beta functions are found equating the $\mu $-derivative of $\mathcal{L}_{%
\mathrm{B}}$ to zero and performing a number of straightforward
manipulations. We find 
\begin{eqnarray}
\beta _{g}^{ijkm} &=&-\frac{1}{36\pi ^{2}}\sum_{np}t_{np}\left[
6(1-s_{np})g_{ijnp}(g_{pnkm}-g_{pmkn})-2s_{np}g_{inpj}g_{kpnm}\right. 
\nonumber \\
&&+2s_{np}(g_{inkp}g_{pjnm}+16f_{jpmn}f_{inkp}^{*})-(3+s_{np})(g_{inpm}g_{kpnj}-g_{inkp}g_{pmnj}-16f_{mpjn}f_{inkp}^{*})
\nonumber \\
&&+48\left.
(1+s_{np})(f_{jmnp}f_{iknp}^{*}-f_{mpjn}f_{iknp}^{*}-f_{jmnp}f_{kpin}^{*})%
\right] _{\scriptstyle (i,j)\leftrightarrow (k,m)},  \label{betas} \\
\beta _{f}^{ijkm} &=&-\frac{1}{18\pi ^{2}}\sum_{np}t_{np}\left[
2(3+s_{np})g_{nipk}f_{mpjn}+3(1+s_{np})g_{nipj}f_{kmnp}+4s_{np}g_{nipk}f_{jpmn}\right] _{%
\scriptstyle (i,j)\leftrightarrow (k,m)}^{\scriptstyle i\leftrightarrow j,\
k\leftrightarrow m},  \nonumber
\end{eqnarray}
and $\beta _{b_{i}}=0$, where $s_{np}=b_{n}b_{p}/|b_{n}b_{p}|$ and $%
t_{np}=1/(|b_{n}|+|b_{p}|)$. The expressions on the right-hand sides of (\ref
{betas}) have to be symmetrized as follows: 
\begin{eqnarray*}
\lbrack X_{ijkm}]_{\scriptstyle (i,j)\leftrightarrow (k,m)} &\equiv &\frac{1%
}{2}\left( X_{ijkm}+X_{kmij}\right) , \\
\lbrack X_{ijkm}]^{\scriptstyle i\leftrightarrow j,\ k\leftrightarrow m}
&\equiv &\frac{1}{4}\left( X_{ijkm}+X_{jikm}+X_{ijmk}+X_{jimk}\right) .
\end{eqnarray*}

\section{Explicit examples}

\setcounter{equation}{0}

In this section we consider some particular cases in detail. A separate
section is devoted to the conditions for asymptotic freedom.

\paragraph{$U(N_{L})$ model}

A simple model is the $U(N_{L})$-symmetric model of left-handed fermions $L$%
, 
\[
\mathcal{L}_{\mathrm{L}}=L^{\dagger i}i\left( \hat{\partial}+b\mathbf{\sigma 
}\cdot \mathbf{\bar{\partial}}\frac{{\bar{\partial}}^{2}}{\Lambda _{L}^{2}}%
\right) L^{i}+\frac{g_{1}}{2\Lambda _{L}^{2}}(L^{\dagger i}L^{i})^{2}+\frac{%
g_{2}}{2\Lambda _{L}^{2}}(L^{\dagger i}L^{j})(L^{\dagger j}L^{i}). 
\]
Here we can set $b=1$ rescaling the space coordinates, the fields and the
couplings. We have a model (\ref{4f}) with 
\[
g_{ijkm}=g_{1}\delta _{ij}\delta _{km}+g_{2}\delta _{im}\delta _{kj},\qquad
f_{ijkm}=0, 
\]
so the one-loop beta functions read 
\begin{equation}
\beta _{1}=\frac{g_{2}^{2}}{36\pi ^{2}}(N_{L}-2),\qquad \beta _{2}=\frac{%
g_{2}^{2}}{36\pi ^{2}}(2N_{L}-1).  \label{bi2}
\end{equation}

The solutions of the RG equations read 
\[
g_{1}(t)=g_{1}(0)+\frac{N_{L}-2}{2N_{L}-1}\left( g_{2}(t)-g_{2}(0)\right)
,\qquad g_{2}(t)=\frac{g_{2}(0)}{1-(2N_{L}-1)g_{2}(0)t/(36\pi ^{2})}, 
\]
where $t=-\ln (|x|\mu )$ and $x$ is some scale.

In the ultraviolet limit ($t\rightarrow \infty $) we have 
\[
g_{1}(t)\sim g_{1}(0)-\frac{N_{L}-2}{2N_{L}-1}g_{2}(0)-\frac{36\pi
^{2}(N_{L}-2)}{(2N_{L}-1)^{2}t},\qquad g_{2}(t)\sim -\frac{36\pi ^{2}}{%
(2N_{L}-1)t}. 
\]
In particular, 
\[
g_{1}(\infty )\sim g_{1}(0)-\frac{N_{L}-2}{2N_{L}-1}g_{2}(0). 
\]
If $g_{1}(\infty )\neq 0$ the UV fixed point is interacting, if $%
g_{1}(\infty )=0$ it is free. However, an interacting fixed point is not
guaranteed to survive beyond the one-loop approximation. Thus, we must
restrict to the subspace with $g_{1}(\infty )=0$.

It is easy to prove that the model with $g_{2}\equiv 0$ is renormalizable,
for example introducing an auxiliary field $\chi $ of weight 1 and writing
the Lagrangian in the form 
\begin{equation}
\mathcal{L}_{\mathrm{L}}^{\prime }=L^{\dagger i}i\left( \hat{\partial}+b%
\mathbf{\sigma }\cdot \mathbf{\bar{\partial}}\frac{{\bar{\partial}}^{2}}{%
\Lambda _{L}^{2}}\right) L^{i}+\chi (L^{\dagger i}L^{i})-\frac{\Lambda
_{L}^{2}}{2g_{1}}\chi ^{2}.  \label{zuto}
\end{equation}
Such a model has vanishing one-loop beta function. However, it is unlikely
to be finite, since no symmetry appears to forbid higher-order corrections.

We have just pointed out a general feature of our four fermion models: there
exist combinations of couplings that have zero one-loop beta functions.
Nevertheless, we are unable to use this observation to prove the existence
of interacting fixed points. That would require more knowledge about
higher-order corrections. Using only one-loop results the best we can do is
to project onto a suitable subspace of the space of couplings, using a
method inspired by Zimmermann's ``reduction of couplings'' \cite{zimme}, and
study the conditions for asymptotic freedom in that subspace. We first prove
that there exists an analytic solution of the RG equations of the form 
\begin{equation}
g_{1}=c_{1}g_{2}+\sum_{i=2}^{\infty }c_{i}g_{2}^{i}  \label{zim}
\end{equation}
to all orders in the perturbative expansion. Consistence with the RG
equations gives 
\[
\frac{\mathrm{d}g_{1}}{\mathrm{d}g_{2}}=\frac{\beta _{1}}{\beta _{2}}, 
\]
which in turn uniquely determines all coefficients $c_{i}$'s. We find 
\[
c_{1}=\frac{N_{L}-2}{2N_{L}-1}, 
\]
plus recurrence relations of the form 
\[
c_{i}=P_{i}(c_{j<i}), 
\]
where $P_{i}(c_{j<i})$ are well-defined polynomials depending only on the
coefficients $c_{j}$'s with $j<i$. The Zimmermann solution (\ref{zim})
restricts the two-parameter space to a curve. There our theory has a unique
coupling, $g_{2}$, and its one-loop beta function is still given by the
second formula of (\ref{bi2}). The condition of asymptotic freedom is thus 
\[
g_{2}<0. 
\]

\paragraph{``Electroweak'' model}

Now we consider a four fermion model containing one family of the
electroweak model. We have the left-handed doublet $L^{a}=(\nu _{L},e_{L})$
and the right-handed electron $e_{R}$. The high-energy four fermion
Lagrangian reads 
\begin{eqnarray*}
\mathcal{L}_{\mathrm{EW}} &=&L^{\dagger a}i\left( \hat{\partial}+b_{L}%
\mathbf{\sigma }\cdot \mathbf{\bar{\partial}}\frac{{\bar{\partial}}^{2}}{%
\Lambda _{L}^{2}}\right) L^{a}+e_{R}^{\dagger }i\left( \hat{\partial}-b_{R}%
\mathbf{\sigma }\cdot \mathbf{\bar{\partial}}\frac{{\bar{\partial}}^{2}}{%
\Lambda _{L}^{2}}\right) e_{R}+\frac{\lambda }{\Lambda _{L}^{2}}(L^{\dagger
a}e_{R})(e_{R}^{\dagger }L^{a}) \\
&&+\frac{g_{1L}}{2\Lambda _{L}^{2}}(L^{\dagger a}L^{a})^{2}+\frac{g_{2L}}{%
2\Lambda _{L}^{2}}(L^{\dagger a}L^{b})(L^{\dagger b}L^{a})+\frac{g_{R}}{%
2\Lambda _{L}^{2}}(e_{R}^{\dagger }e_{R})^{2}+\frac{g_{LR}}{\Lambda _{L}^{2}}%
(L^{\dagger a}L^{a})(e_{R}^{\dagger }e_{R}).
\end{eqnarray*}
Because of hypercharge conservation $f$-terms are not allowed, so this model
is of restricted type (\ref{l4fr}). Define $\ell ^{i}=(\nu
_{L},e_{L},e_{R}^{c})$, where $^{c}$ denotes the charge conjugate. Then we
have 
\begin{eqnarray*}
g_{ijkm} &=&g_{1L}\hat{\delta}_{ij}\hat{\delta}_{km}+g_{2L}\hat{\delta}_{im}%
\hat{\delta}_{jk}+g_{R}\delta _{i3}\delta _{j3}\delta _{k3}\delta _{m3} \\
&&+(\lambda -g_{LR})(\hat{\delta}_{ij}\delta _{k3}\delta _{m3}+\delta
_{i3}\delta _{j3}\hat{\delta}_{km})+\lambda (\hat{\delta}_{im}\delta
_{k3}\delta _{j3}+\delta _{i3}\delta _{m3}\hat{\delta}_{jk}),
\end{eqnarray*}
where $\hat{\delta}_{ij}=\delta _{ij}-\delta _{i3}\delta _{j3}$. Applying (%
\ref{betas}) we obtain the beta functions 
\begin{eqnarray}
\beta _{1L} &=&\frac{\lambda ^{2}}{36\pi ^{2}|b_{R}|},\qquad \beta _{2L}=%
\frac{1}{36\pi ^{2}}\left( 3\frac{g_{2L}^{2}}{|b_{L}|}+2\frac{\lambda ^{2}}{%
|b_{R}|}\right) ,\qquad \beta _{R}=\frac{\lambda ^{2}}{6\pi ^{2}|b_{L}|}, 
\nonumber \\
\qquad \beta _{LR} &=&\frac{\lambda }{36\pi ^{2}}\left( u+\frac{\lambda (s-3)%
}{|b_{L}|+|b_{R}|}\right) ,\qquad \beta _{\lambda }=\frac{\lambda }{18\pi
^{2}}\left( u-\frac{\lambda (2s+3)}{|b_{L}|+|b_{R}|}\right) ,\qquad
\label{ba}
\end{eqnarray}
where $s\equiv b_{L}b_{R}/(|b_{L}||b_{R}|)$ and 
\[
u\equiv \frac{g_{1L}}{|b_{L}|}+2\frac{g_{2L}}{|b_{L}|}+\frac{g_{R}}{|b_{R}|}+%
\frac{4sg_{LR}}{|b_{L}|+|b_{R}|}. 
\]
Recall that $b_{L}$ and $b_{R}$ do not run at one loop.

We see that the beta functions depend only on three couplings, precisely $%
\lambda $, $g_{2L}$ and the combination $u$. The conditions for asymptotic
freedom are studied in section 5. There exist RG-flow trajectories with
special properties, possibly hidden symmetries. This topic is discussed in
section 6.

\paragraph{$U(N_{L})\times U(N_{R})$ model}

This is a generalization of the electroweak model, where the left- and
right-handed fermions are in the fundamental representations of $U(N_{L})$
and $U(N_{R})$, respectively. The Lagrangian is 
\begin{eqnarray*}
\mathcal{L}_{N_{L}N_{R}} &=&L^{\dagger a}i\left( \hat{\partial}+b_{L}\mathbf{%
\sigma }\cdot \mathbf{\bar{\partial}}\frac{{\bar{\partial}}^{2}}{\Lambda
_{L}^{2}}\right) L^{a}+E_{R}^{\dagger I}i\left( \hat{\partial}-b_{R}\mathbf{%
\sigma }\cdot \mathbf{\bar{\partial}}\frac{{\bar{\partial}}^{2}}{\Lambda
_{L}^{2}}\right) E_{R}^{I} \\
&&+\frac{g_{1L}}{2\Lambda _{L}^{2}}(L^{\dagger a}L^{a})^{2}+\frac{g_{2L}}{%
2\Lambda _{L}^{2}}(L^{\dagger a}L^{b})(L^{\dagger b}L^{a})+\frac{g_{1R}}{%
2\Lambda _{L}^{2}}(E_{R}^{\dagger I}E_{R}^{I})^{2}+\frac{g_{2R}}{2\Lambda
_{L}^{2}}(E_{R}^{\dagger I}E_{R}^{J})(E_{R}^{\dagger J}E_{R}^{I}) \\
&&+\frac{g_{1LR}}{\Lambda _{L}^{2}}(L^{\dagger a}L^{a})(E_{R}^{\dagger
I}E_{R}^{I})+\frac{g_{2LR}}{\Lambda _{L}^{2}}(L^{\dagger
a}E_{R}^{I})(E_{R}^{\dagger I}L^{a}),
\end{eqnarray*}
where $a,b=1,\ldots N_{L}$ and $I,J=1,\ldots N_{R}$. Define $\ell
^{i}=(L^{a},E^{cI})$, with $i=(a,I)$. We have a restricted model (\ref{l4fr}%
) with couplings 
\begin{eqnarray*}
g_{ijkm} &=&g_{1L}\hat{\delta}_{ij}\hat{\delta}_{km}+g_{2L}\hat{\delta}_{im}%
\hat{\delta}_{jk}+g_{1R}\bar{\delta}_{ij}\bar{\delta}_{km}+g_{2R}\bar{\delta}%
_{im}\bar{\delta}_{jk} \\
&&+(g_{2LR}-g_{1LR})(\hat{\delta}_{ij}\bar{\delta}_{km}+\hat{\delta}_{km}%
\bar{\delta}_{ij})+g_{2LR}(\hat{\delta}_{im}\bar{\delta}_{jk}+\hat{\delta}%
_{jk}\bar{\delta}_{im}),
\end{eqnarray*}
where $\hat{\delta}$ and $\bar{\delta}$ are the Kronecker tensors of $%
U(N_{L})$ and $U(N_{R})$, respectively. The beta functions are 
\begin{eqnarray}
\beta _{1L} &=&\frac{1}{36\pi ^{2}}\left( (N_{L}-2)\frac{g_{2L}^{2}}{|b_{L}|}%
+N_{R}\frac{g_{2LR}^{2}}{|b_{R}|}\right) ,\qquad \beta _{2L}=\frac{1}{36\pi
^{2}}\left( (2N_{L}-1)\frac{g_{2L}^{2}}{|b_{L}|}+2N_{R}\frac{g_{2LR}^{2}}{%
|b_{R}|})\right) ,  \nonumber \\
\beta _{1R} &=&\frac{1}{36\pi ^{2}}\left( (N_{R}-2)\frac{g_{2R}^{2}}{|b_{R}|}%
+N_{L}\frac{g_{2LR}^{2}}{|b_{L}|}\right) ,\qquad \beta _{2R}=\frac{1}{36\pi
^{2}}\left( (2N_{R}-1)\frac{g_{2R}^{2}}{|b_{R}|}+2N_{L}\frac{g_{2LR}^{2}}{%
|b_{L}|})\right) ,  \nonumber \\
\beta _{1LR} &=&\frac{g_{2LR}}{36\pi ^{2}}\left( u+\frac{g_{2LR}(s-3)}{%
|b_{L}|+|b_{R}|}\right) ,\qquad \qquad \qquad \beta _{2LR}=\frac{g_{2LR}}{%
18\pi ^{2}}\left( u-\frac{g_{2LR}(2s+3)}{|b_{L}|+|b_{R}|}\right) ,
\label{lr}
\end{eqnarray}
where now 
\[
u=\frac{g_{1L}}{|b_{L}|}+\frac{g_{1R}}{|b_{R}|}+N_{L}\frac{g_{2L}}{|b_{L}|}%
+N_{R}\frac{g_{2R}}{|b_{R}|}+\frac{4sg_{1LR}}{|b_{L}|+|b_{R}|}.
\]

\paragraph{Dirac fermions}

So far, we have considered only explicit examples of reduced type (\ref{l4fr}%
). Now we consider a model of $N$ Dirac fermions, which involves also $f$
terms and new types of $g$ terms. We impose the flavor symmetry $U(N)$ and
parity invariance, which allows us to set $b_{L}=b_{R}=1$.

As usual, write the Dirac fermions $\psi ^{i}$ as $(\ell _{1}^{i},\ell
_{2}^{ci})$, where $\ell _{1}^{i}$ denote the left-handed components and $%
\ell _{2}^{i}$ are the charge-conjugates of the right-handed components. The
action of parity reads 
\[
P\ell _{1}^{i}=\varepsilon \ell _{2}^{i*},\qquad P\ell _{2}^{i}=-\varepsilon
\ell _{1}^{i*}. 
\]
Moreover, $\ell _{1}^{i}$ and $\ell _{2}^{i}$ belong to the fundamental and
anti-fundamental $U(N)$ representations, respectively.

The Lagrangian (\ref{4f})\ becomes 
\begin{eqnarray}
\mathcal{L}_{N} &=&\ell _{1}^{\dagger i}i\left( \hat{\partial}+\mathbf{%
\sigma }\cdot \mathbf{\bar{\partial}}\frac{{\bar{\partial}}^{2}}{\Lambda
_{L}^{2}}\right) \ell _{1}^{i}+\ell _{2}^{\dagger i}i\left( \hat{\partial}+%
\mathbf{\sigma }\cdot \mathbf{\bar{\partial}}\frac{{\bar{\partial}}^{2}}{%
\Lambda _{L}^{2}}\right) \ell _{2}^{i}  \nonumber \\
&&+\frac{g_{1}}{2\Lambda _{L}^{2}}\left[ (\ell _{1}^{\dagger i}\ell
_{1}^{i})^{2}+(\ell _{2}^{\dagger i}\ell _{2}^{i})^{2}\right] +\frac{g_{2}}{%
2\Lambda _{L}^{2}}\left[ (\ell _{1}^{\dagger i}\ell _{1}^{j})(\ell
_{1}^{\dagger j}\ell _{1}^{i})+(\ell _{2}^{\dagger i}\ell _{2}^{j})(\ell
_{2}^{\dagger j}\ell _{2}^{i})\right]  \label{diracco} \\
&&+\frac{g_{3}}{\Lambda _{L}^{2}}(\ell _{1}^{\dagger i}\ell _{1}^{i})(\ell
_{2}^{\dagger j}\ell _{2}^{j})+\frac{g_{4}}{\Lambda _{L}^{2}}(\ell
_{1}^{\dagger i}\ell _{2}^{j})(\ell _{2}^{\dagger j}\ell _{1}^{i})+\frac{%
g_{5}}{\Lambda _{L}^{2}}(\ell _{1}^{\dagger i}\ell _{1}^{j})(\ell
_{2}^{\dagger i}\ell _{2}^{j})+\frac{g_{6}}{\Lambda _{L}^{2}}(\ell
_{1}^{\dagger i}\ell _{2}^{j})(\ell _{2}^{\dagger i}\ell _{1}^{j})  \nonumber
\\
&&+\frac{f_{1}}{\Lambda _{L}^{2}}\left[ (\ell _{1}^{Ti}\varepsilon \ell
_{2}^{i})(\ell _{1}^{Tj}\varepsilon \ell _{2}^{j})+(\ell _{1}^{\dagger
i}\varepsilon \ell _{2}^{*i})(\ell _{1}^{\dagger j}\varepsilon \ell
_{2}^{*j})\right] +\frac{f_{2}}{\Lambda _{L}^{2}}\left[ (\ell
_{1}^{Ti}\varepsilon \ell _{2}^{j})(\ell _{1}^{Tj}\varepsilon \ell
_{2}^{i})+(\ell _{1}^{\dagger i}\varepsilon \ell _{2}^{*j})(\ell
_{1}^{\dagger j}\varepsilon \ell _{2}^{*i})\right] ,  \nonumber
\end{eqnarray}
and all couplings are real. A possible third vertex of $f$-type, namely 
\[
(\ell _{1}^{Ti}\varepsilon \ell _{1}^{j})(\ell _{2}^{Ti}\varepsilon \ell
_{2}^{j}) 
\]
is not included, because it is not independent of the other two. Indeed, a
Fierz rearrangement gives the identity 
\[
(\ell _{1}^{Ti}\varepsilon \ell _{2}^{i})(\ell _{1}^{Tj}\varepsilon \ell
_{2}^{j})+(\ell _{1}^{Ti}\varepsilon \ell _{2}^{j})(\ell
_{1}^{Tj}\varepsilon \ell _{2}^{i})+(\ell _{1}^{Ti}\varepsilon \ell
_{1}^{j})(\ell _{2}^{Ti}\varepsilon \ell _{2}^{j})=0. 
\]

We report only the beta functions in the large $N$ limit, which simplify
considerably: 
\begin{eqnarray*}
\beta _{g_{1}} &=&\frac{N(g_{2}^{2}+g_{4}^{2})}{36\pi ^{2}},\qquad \beta
_{g_{2}}=\frac{N(g_{2}^{2}+g_{4}^{2})}{18\pi ^{2}},\qquad \beta _{g_{3}}=%
\frac{Ng_{2}g_{4}}{18\pi ^{2}},\qquad \beta _{g4}=\frac{Ng_{2}g_{4}}{9\pi
^{2}}, \\
\beta _{g_{5}} &=&-\frac{N}{18\pi ^{2}}\left(
g_{5}^{2}+g_{5}g_{6}+g_{6}^{2}+12f_{1}^{2}-12f_{1}f_{2}+4f_{2}^{2}\right) ,
\\
\beta _{g_{6}} &=&-\frac{N}{36\pi ^{2}}\left(
g_{5}^{2}+4g_{5}g_{6}+g_{6}^{2}+24f_{1}^{2}-24f_{1}f_{2}+4f_{2}^{2}\right) ,
\\
\beta _{f_{1}} &=&\frac{N}{18\pi ^{2}}\left[
f_{2}(g_{5}+2g_{6})-3f_{1}(g_{5}+g_{6})\right] ,\qquad \beta _{f_{2}}=\frac{N%
}{18\pi ^{2}}f_{2}(g_{6}-g_{5}).
\end{eqnarray*}
Observe that the beta functions do not depend on $g_{1}$ and $g_{3}$.
Moreover, the couplings separate in two groups, $g_{1\text{-}4}$ and $%
g_{5,6} $-$f_{1,2}$, which do not talk to each other. These, however, are
only features of the large $N$ limit.

\section{Asymptotic freedom}

\setcounter{equation}{0}

In this section we study the conditions for asymptotic freedom in the
presence of more than one coupling. The idea is to search for solutions of
the RG\ equations as expansions around the ultraviolet free fixed point. The
domain of asymptotic freedom $\mathcal{D}_{\text{AF}}$ is then determined by
the free parameters contained in the solution. A different approach to
asymptotic freedom with more couplings is due to Zimmermann \cite{zimme2}.
An investigation that is in part related to this problem can be found in 
\cite{oehme}.

We first illustrate our method in the case of a single coupling $\alpha $
with beta function 
\[
\beta _{\alpha }=\dot{\alpha}=\alpha \sum_{n=1}^{\infty }\beta _{n}\alpha
^{n}.
\]
where the dot denotes the derivative with respect to $t=-\ln (|x|\mu )$, $|x|
$ being some scale. If $\beta _{1}\neq 0$ the asymptotic expansion around
the ultraviolet limit $t\rightarrow \infty $ reads 
\begin{equation}
\alpha (t)=\frac{1}{t}\sum_{n=0}^{\infty }\frac{b_{n}(\ln t)}{t^{n}},
\label{as1}
\end{equation}
where $b_{n}$ are polynomials of degree $n$ in $\ln t$. Inserting this
expansion into the RG equation we get $b_{0}=-1/\beta _{1}$ and the
recursion relations 
\begin{equation}
(n-1)b_{n}-b_{n}^{\prime }=\delta _{n<}  \label{recu}
\end{equation}
for $n>0$, where $\delta _{n<}$ is a linear combination of monomials $%
\prod_{i}b_{n_{i}}^{k_{i}}$ with $\sum_{i}n_{i}k_{i}\leqslant n$ and depends
only on the coefficients $b_{m}$ with $0<m<n$. Consider first $n=1$, and
observe that $\delta _{1<}$ contains no logarithms, so $b_{1}$ is a
polynomial of degree 1. The coefficient of $\ln t$ in $b_{1}$ is uniquely
determined, while $b_{1}(0)\equiv b$ remains arbitrary. For $n>1$ the
relations (\ref{recu}) can be solved recursively: 
\begin{equation}
b_{n}=\frac{1}{n-1}(\delta _{n<}+b_{n}^{\prime })=\frac{1}{n-1}\delta _{n<}+%
\frac{1}{(n-1)^{2}}\delta _{n<}^{\prime }+\frac{1}{(n-1)^{3}}\delta
_{n<}^{\prime \prime }+\cdots   \label{simi}
\end{equation}
Clearly, the sum ends after a finite number of terms, since $\delta _{n<}$
is a polynomial.

Thus, the asymptotic solution (\ref{as1}) is well-defined and uniquely
determined as a function of the arbitrary constant $b$. To the lowest
orders, we find 
\begin{equation}
\alpha (t)=-\frac{1}{\beta _{1}t}-\frac{\beta _{2}}{\beta _{1}^{3}}\frac{\ln
t}{t^{2}}+\frac{b}{t^{2}}-\frac{\beta _{2}^{2}}{\beta _{1}^{5}}\frac{\ln
^{2}t}{t^{3}}+\frac{\beta _{2}}{\beta _{1}^{5}}(2b\beta _{1}^{3}+\beta _{2})%
\frac{\ln t}{t^{3}}+\mathcal{O}\left( t^{-3}\right) .  \label{2log}
\end{equation}
If $\beta _{1}=0$ but $\beta _{2}<0$ we have the expansion 
\begin{equation}
\alpha (t)=\frac{1}{t^{1/2}}\sum_{n=0}^{\infty }\frac{b_{n}(\ln t)}{t^{n/2}}=%
\frac{1}{\sqrt{-2\beta _{2}t}}+\frac{\beta _{3}}{2\beta _{2}^{2}t}+\frac{%
\beta _{3}^{2}-\beta _{2}\beta _{4}}{4\sqrt{2}(-\beta _{2})^{7/2}}\frac{\ln t%
}{t^{3/2}}+\frac{b}{t^{3/2}}+\mathcal{O}\left( t^{-2}\ln t\right) .
\label{b2}
\end{equation}
If the first non-vanishing coefficient is $\beta _{n}$ then the expansion
begins with $t^{-1/n}$.

Now we generalize this result to the case of more couplings. Consider a
theory with $s$ couplings $g=\{g_{i}\}$ and beta functions $\beta _{i}(g)$.
To be specific, we assume 
\begin{equation}
\beta _{i}(g)=\sum_{n=2}^{\infty }\frac{1}{n!}c_{ij_{1}\cdots
j_{n}}g_{j_{1}}\cdots g_{j_{n}}=\frac{1}{2}c_{ijk}g_{j}g_{k}+\mathcal{O}%
(g^{3}),  \label{beta}
\end{equation}
where the constants $c_{ijk}$ are the one-loop coefficients. We look for
asymptotic solutions of the RG equations starting form the ``Zimmermann
trajectories'' 
\begin{equation}
g_{i}(t)\sim -\frac{a_{i}}{t}  \label{expa}
\end{equation}
in the limit $t\rightarrow \infty $. Inserting (\ref{expa}) into the RG\
equations 
\begin{equation}
\dot{g}_{i}(t)=\beta _{i}(g(t)),  \label{rgeq}
\end{equation}
where $\beta _{i}$ are given by (\ref{beta}), and keeping only the leading
terms, we see that the constants $a_{i}$ are determined by the quadratic
equations 
\begin{equation}
a_{i}=\frac{1}{2}c_{ijk}a_{j}a_{k},  \label{ztr}
\end{equation}
which have, in general, a discrete set of solutions. Normally the solutions
just have to be real (if the couplings are parametrized to be real, as we
assume), but in some cases further physical restrictions might apply. For
example, stability (positive-definiteness of the action in Euclidean space)
might require that some couplings be positive. We do not consider such
restrictions here and assume that all real solutions are physical
acceptable. It is straightforward to adapt our conclusions to specific
situations.

Around a Zimmermann trajectory, we continue the expansion as 
\begin{equation}
g_{i}(t)\sim -\frac{1}{t}\left( a_{i}+\frac{b_{i}}{t^{\gamma }}\right) ,
\label{star}
\end{equation}
assuming that $\gamma $ is a positive number. Then $\gamma $ and $b_{i}$ are
an eigenvalue and an eigenvector of the real matrix 
\[
\mathcal{N}_{ij}=c_{ikj}a_{k}-\delta _{ij}, 
\]
respectively.

The matrix $\mathcal{N}$ is crucial for our discussion. The dimension $d_{%
\text{AF}}$ of the domain $\mathcal{D}_{\text{AF}}$ of asymptotic freedom is
equal to the number of $\mathcal{N}$-positive eigenvalues $\gamma $ ,
including multiplicities. Observe that because of (\ref{ztr}) one eigenvalue
is always equal to 1, with eigenvector $a_{i}$. If the Zimmermann trajectory
exists, the dimension $d_{\text{AF}}$ is at least $1$. If $\gamma =1$ is the
unique positive eigenvalue, the form of the expansion is (\ref{as1}).

The most general solution reads 
\begin{equation}
g_{i}(t)=-\frac{1}{t}\left( a_{i}+\sum_{\mathbf{\gamma }\cdot \mathbf{n},n>0}%
\frac{b_{i,\mathbf{\gamma }\cdot \mathbf{n}}(\ln t)}{t^{\mathbf{\gamma }%
\cdot \mathbf{n}}}\right) .  \label{ag2}
\end{equation}
Here $\mathbf{\gamma }$ is a vector collecting the positive eigenvalues of
the matrix $\mathcal{N}$, while $\mathbf{n}$ is a vector of non-negative
integer entries. The condition $n>0$ means that $\mathbf{n}$ must not vanish
identically. Two vectors $\mathbf{n}$ and $\mathbf{n}^{\prime }$ such that $%
\mathbf{\gamma }\cdot \mathbf{n=\gamma }\cdot \mathbf{n}^{\prime }$ are
equivalent, and associated with a unique numerator $b_{i,\mathbf{\gamma }%
\cdot \mathbf{n}}$. The sum is ordered for increasing values of $\mathbf{%
\gamma }\cdot \mathbf{n}$. Finally, the $b_{i,\mathbf{\gamma }\cdot \mathbf{n%
}}(\ln t)$'s are polynomials of certain finite degrees in $\ln t$.

Inserting (\ref{ag2}) into the RG\ equations (\ref{rgeq}) and isolating the
coefficients of the powers $t^{-2+\mathbf{\gamma }\cdot \mathbf{n}}$, we
obtain equations for the polynomials $b_{i,\mathbf{\gamma }\cdot \mathbf{n}}$%
. It is immediate to find that such equations have the form 
\begin{equation}
\left[ (\mathbf{\gamma }\cdot \mathbf{n})\delta _{ij}-\mathcal{N}%
_{ij}\right] b_{j,\mathbf{\gamma }\cdot \mathbf{n}}-b_{i,\mathbf{\gamma }%
\cdot \mathbf{n}}^{\prime }=\delta _{i,\mathbf{\gamma }\cdot \mathbf{n<}},
\label{equus}
\end{equation}
where $\delta _{i,\mathbf{\gamma }\cdot \mathbf{n<}}$ is a sum of monomials 
\[
\prod_{k}b_{j,\mathbf{\gamma }\cdot \mathbf{n}_{k}} 
\]
with $\mathbf{\gamma }\cdot \mathbf{n}_{k}<\mathbf{\gamma }\cdot \mathbf{n}$
and $\sum_{k}\mathbf{\gamma }\cdot \mathbf{n}_{k}\leqslant \mathbf{\gamma }%
\cdot \mathbf{n}$.

Clearly, $\delta _{i,\mathbf{\gamma }\cdot \mathbf{n<}}$ contains a finite
number of terms. Now we want to show that equations (\ref{equus}) allow us
to recursively determine the $b_{i,\mathbf{\gamma }\cdot \mathbf{n}}$'s.

For intermediate purposes, it is convenient to turn to the basis where the
matrix $\mathcal{N}_{ij}$ has a real canonical Jordan form. Quantities in
this basis are denoted with a tilde. Specifically, $\mathcal{\tilde{N}}$ is
block-diagonal. Its first blocks are associated with the real eigenvalues $%
\lambda $ and have the form 
\begin{equation}
\left( 
\begin{tabular}{ccc}
$\lambda $ & $0$ & $0$ \\ 
$1$ & $\ddots $ & $0$ \\ 
$0$ & $1$ & $\lambda $%
\end{tabular}
\right) ,  \label{sta}
\end{equation}
while the last blocks are associated with the complex eigenvalues $\mu $ and
have the same form as (\ref{sta}), where however the $\lambda $'s are
replaced by 2$\times $2 blocks 
\[
\left( 
\begin{tabular}{cc}
Re$\mu $ & Im$\mu $ \\ 
$-$Im$\mu $ & Re$\mu $%
\end{tabular}
\right) , 
\]
the 1's are replaced by 2$\times $2 identity matrices and the 0's are
replaced by 2$\times $2 matrices with vanishing entries. All matrices $%
\mathcal{\tilde{N}}_{ij}-(\mathbf{\gamma }\cdot \mathbf{n})\delta _{ij}$ are
then in canonical Jordan form.

Let $M_{ij}$ be such that $\mathcal{N}=M^{-1}\mathcal{\tilde{N}}M$ and $%
\tilde{b}_{i,\mathbf{\gamma }\cdot \mathbf{n}}\equiv M_{ij}b_{j,\mathbf{%
\gamma }\cdot \mathbf{n}}$. Multiplying equation (\ref{equus}) by $M$ to the
left, we can rewrite it in the form 
\begin{equation}
\left[ (\mathbf{\gamma }\cdot \mathbf{n})\delta _{ij}-\mathcal{\tilde{N}}%
_{ij}\right] \tilde{b}_{j,\mathbf{\gamma }\cdot \mathbf{n}}-\tilde{b}_{i,%
\mathbf{\gamma }\cdot \mathbf{n}}^{\prime }=\tilde{\delta}_{i,\mathbf{\gamma 
}\cdot \mathbf{n<}},  \label{equus2}
\end{equation}
By induction, if we assume that the polynomials $\tilde{b}_{i,\mathbf{\gamma 
}\cdot \mathbf{n}^{\prime }}$ with $\mathbf{\gamma }\cdot \mathbf{n}^{\prime
}<\mathbf{\gamma }\cdot \mathbf{n}$ are known, we conclude that the $\tilde{%
\delta}_{i,\mathbf{\gamma }\cdot \mathbf{n<}}$'s are polynomials of certain
finite degrees in $\ln t$. When the matrix $\mathcal{\tilde{N}}_{ij}-(%
\mathbf{\gamma }\cdot \mathbf{n})\delta _{ij}$ is invertible, the
polynomials $\tilde{b}_{i,\mathbf{\gamma }\cdot \mathbf{n}}$ are uniquely
determined. Calling $\mathcal{U}_{ij}$ the inverse matrix of $\mathcal{%
\tilde{N}}_{ij}-(\mathbf{\gamma }\cdot \mathbf{n})\delta _{ij}$, we have,
similarly to (\ref{simi}), 
\begin{equation}
\tilde{b}_{i,\mathbf{\gamma }\cdot \mathbf{n}}=-\mathcal{U}_{ij}\left( 
\tilde{\delta}_{j,\mathbf{\gamma }\cdot \mathbf{n<}}+\tilde{b}_{j,\mathbf{%
\gamma }\cdot \mathbf{n}}^{\prime }\right) =-\mathcal{U}_{ij}\tilde{\delta}%
_{j,\mathbf{\gamma }\cdot \mathbf{n<}}+\mathcal{U}_{ij}\mathcal{U}_{jk}%
\tilde{\delta}_{k,\mathbf{\gamma }\cdot \mathbf{n<}}^{\prime }-\mathcal{U}%
_{ij}\mathcal{U}_{jk}\mathcal{U}_{kl}\tilde{\delta}_{l,\mathbf{\gamma }\cdot 
\mathbf{n<}}^{\prime \prime }+\cdots  \label{gre}
\end{equation}
Again, the sum ends after a finite number of terms, since the $\tilde{\delta}%
_{i,\mathbf{\gamma }\cdot \mathbf{n<}}$'s are polynomials.

When $\mathcal{\tilde{N}}_{ij}-(\mathbf{\gamma }\cdot \mathbf{n})\delta _{ij}
$ is not invertible, one of its blocks has $m_{0}>0$ zeros on the diagonal.
Assume that this block is the one with $\bar{\imath}\leqslant i,j<\bar{\imath%
}+m_{0}$ and proceed as follows. The block-structure of $\mathcal{\tilde{N}}$
allows us to split equation (\ref{equus2}) into: $a$)\ the equation for the $%
\tilde{b}_{i,\mathbf{\gamma }\cdot \mathbf{n}}$'s with $i<\bar{\imath}$; $b$%
) the equation for the $\tilde{b}_{i,\mathbf{\gamma }\cdot \mathbf{n}}$'
with $\bar{\imath}\leqslant i,j<\bar{\imath}+m_{0}$; $c$) the equation for
the $\tilde{b}_{i,\mathbf{\gamma }\cdot \mathbf{n}}$'s with $i\geqslant \bar{%
\imath}+m_{0}$. Equations \noindent $a$) and $c$) are solved by formulas
similar to (\ref{gre}). Equation \noindent $b$) has the form 
\[
\left( 
\begin{tabular}{ccc}
$0$ & $0$ & $0$ \\ 
$\zeta _{1}$ & $\ddots $ & $0$ \\ 
$0$ & $\zeta _{m_{0}-1}$ & $0$%
\end{tabular}
\right) _{ij}\tilde{b}_{j,\mathbf{\gamma }\cdot \mathbf{n}}-\tilde{b}_{i,%
\mathbf{\gamma }\cdot \mathbf{n}}^{\prime }=\tilde{\delta}_{i,\mathbf{\gamma 
}\cdot \mathbf{n<}},
\]
where the $\zeta _{i}$'s can be equal to 0 or 1. The right-hand side is made
of recursively known polynomials of some degrees $d_{i}$ in $\ln t$. Then
the $\tilde{b}_{i,\mathbf{\gamma }\cdot \mathbf{n}}$'s are polynomials of
finite degrees greater than $d_{i}$, and each of them is uniquely determined
up to an arbitrary additional constant. Therefore, in total we have $m_{0}$
arbitrary constants.

Thus, equations (\ref{equus})-(\ref{equus2}) can be solved recursively to
determine the polynomials $\tilde{b}_{i,\mathbf{\gamma }\cdot \mathbf{n}}$,
and therefore the $b_{i,\mathbf{\gamma }\cdot \mathbf{n}}$'s. The solutions (%
\ref{ag2}) contain a number of arbitrary constants equal to the number of
times the matrices $\mathcal{\tilde{N}}_{ij}-(\mathbf{\gamma }\cdot \mathbf{n%
})\delta _{ij}$ become degenerate, including multiplicities. This number is
equal to the number of positive $\mathcal{N}$-eigenvalues, including
multiplicities. Indeed, recalling that different $\mathbf{n}$'s with the
same $\mathbf{\gamma }\cdot \mathbf{n}$ are equivalent, each equation $%
\gamma _{i}=\mathbf{\gamma }\cdot \mathbf{n}$ admits precisely one solution,
and no other degeneracies are possible.

The set of arbitrary constants contained in the asymptotic expansion (\ref
{ag2}) determines the domain of asymptotic freedom $\mathcal{D}_{\text{AF}}$%
. We conclude that the dimension of $\mathcal{D}_{\text{AF}}$ is equal to
the number of positive eigenvalues $\gamma $ of the matrix $\mathcal{N}$,
including multiplicities.

In practice, we have to look for the Zimmermann trajectory around which the
asymptotic expansion (\ref{ag2}) has the maximal number of positive
eigenvalues. In most cases the other Zimmermann trajectories also play a
role. They can determine the boundary of $\mathcal{D}_{\text{AF}}$, if it is
two-dimensional, or its edges, if it more than two dimensional.

Now we consider two examples in detail.

\paragraph{Dirac model}

First we consider the Dirac model (\ref{diracco}) in the large $N$ limit,
focusing on the $g_{2}$-$g_{4}$ subsystem. The RG flow is given by the
equations 
\[
\beta _{2}=\dot{g}_{2}=\kappa (g_{2}^{2}+g_{4}^{2})+\mathcal{O}%
(g^{3}),\qquad \beta _{4}=\dot{g}_{4}=2\kappa g_{2}g_{4}+\mathcal{O}(g^{3}), 
\]
where $\kappa =N/(18\pi ^{2})>0$. We find the constants 
\[
c_{222}=c_{244}=c_{424}=c_{442}=2\kappa , 
\]
while all other entries $c_{ijk}$ vanish. The Zimmermann trajectories are
given by $(a_{2},a_{4})=(1,0)/\kappa $ and $(a_{2},a_{4})=(1,\pm 1)/(2\kappa
)$. Expanding around the trajectories with $(a_{2},a_{4})=(1,0)/\kappa $ we
find that $\mathcal{N}$ is equal to the identity matrix, so $\gamma =1$ with
multiplicity 2. This means that $\mathcal{D}_{\text{AF}}$ has dimension 2.
Two arbitrary constants appear at order $t^{-2}$.

Precisely, we find 
\begin{equation}
g_{2}(t)=-\frac{1}{\kappa t}+\frac{1}{t^{2}}(\xi _{2}\ln t+b_{2})+\mathcal{O}%
\left( t^{-3}\ln ^{2}t\right) ,\qquad g_{4}(t)=\frac{1}{t^{2}}(\xi _{4}\ln
t+b_{4})+\mathcal{O}\left( t^{-3}\ln ^{2}t\right) ,  \label{asyco}
\end{equation}
where $b_{2,4}$ are arbitrary constants, while $\xi _{2,4}$ are uniquely
determined by the cubic terms of the beta functions. Since all other
matrices $\mathcal{N}_{ij}-(\mathbf{\gamma }\cdot \mathbf{n})\delta _{ij}$
are invertible the higher-order corrections are uniquely determined. Thus,
the two-dimensional domain of asymptotic freedom is parametrized by the
arbitrary constants $b_{2,4}$. The other two Zimmermann trajectories, given
by $(a_{2},a_{4})=(1,\pm 1)/(2\kappa )$, are the boundary of $\mathcal{D}_{%
\text{AF}}$.

At one loop we can check our results solving the system explicitly. Call $%
\lambda _{\pm }=g_{2}\pm g_{4}$. We have 
\[
\beta _{\pm }=\dot{\lambda}_{\pm }=\kappa \lambda _{\pm }^{2},
\]
wherefrom 
\begin{equation}
\lambda _{\pm }(t)=\frac{\lambda _{\pm 0}}{1-\kappa \lambda _{\pm 0}t}.
\label{11}
\end{equation}
In the ultraviolet limit $t\rightarrow \infty $ the asymptotic behaviors are 
\[
g_{2}(t)=-\frac{1}{\kappa t}-\frac{1}{2\kappa ^{2}t^{2}}\frac{\lambda
_{+0}+\lambda _{-0}}{\lambda _{+0}\lambda _{-0}}+\mathcal{O}\left(
t^{-3}\right) ,\qquad g_{4}(t)=\frac{1}{2\kappa ^{2}t^{2}}\frac{\lambda
_{+0}-\lambda _{-0}}{\lambda _{+0}\lambda _{-0}}+\mathcal{O}\left(
t^{-3}\right) .
\]
Here the logarithmic terms of (\ref{asyco}) are absent, because we are
neglecting higher-order corrections to the beta functions (check also (\ref
{2log})).

\paragraph{Electroweak model}

Now we apply our method to the electroweak four fermion model (\ref{ba})
with $b_{L}=b_{R}=1$. We first restrict to the subspace $(g_{2L},\lambda ,u)$%
. At one loop we have 
\begin{equation}
\beta _{2L}=\frac{1}{(6\pi )^{2}}(3g_{2L}^{2}+2\lambda ^{2}),\qquad \beta
_{\lambda }=\frac{\lambda }{(6\pi )^{2}}\left( 2u-5\lambda \right) ,\qquad
\beta _{u}=\frac{1}{(6\pi )^{2}}(9\lambda ^{2}+6g_{2L}^{2}+2\lambda u).
\label{ewbet}
\end{equation}
The conditions $\beta _{2L}=\beta _{\lambda }=\beta _{u}=0$ are solved by $%
g_{2L}=\lambda =0$, while $u$ remains arbitrary. Since a non-trivial fixed
point cannot be trusted within our approximation, we need to project onto a
suitable subspace of parameter space. Such a projection is automatic in our
approach.

The Zimmermann trajectories are 
\[
g_{2L}\sim -\frac{a_{1}}{t},\qquad \lambda \sim -\frac{a_{2}}{t},\qquad
u\sim -\frac{a_{3}}{t}, 
\]
with 
\begin{equation}
(a_{1},a_{2},a_{3})=\left( \frac{1}{3},0,\frac{2}{3}\right) ,\left( \frac{1}{%
5},\frac{1}{5},1\right) ,\left( 0.046,-0.141,0.149\right) .  \label{z2}
\end{equation}
The third trajectory is given by complicated irrational coefficients of
which we just give the approximate numerical values. Two complex solutions
to (\ref{ztr}) also exist, but must be discarded.

Expanding around the first Zimmermann trajectory, we find 
\begin{equation}
\mathcal{N}=\left( 
\begin{array}{ccc}
1\  & \ 0\  & \ 0\  \\ 
0\  & \ \frac{1}{3}\  & \ 0\  \\ 
4\  & \ \frac{4}{3}\  & -1
\end{array}
\right) ,  \label{nojordan}
\end{equation}
so the positive eigenvalues are $\gamma =1/3$ and $\gamma =1$. The Jordan
canonical form of this matrix is diagonal: $\mathcal{\tilde{N}}=$diag$%
(1,1/3,-1)$. The domain of asymptotic freedom is two-dimensional and the
arbitrary constants appear at orders $t^{-4/3}$ and $t^{-2}$. Using the
one-loop truncated beta functions, the asymptotic expansion of the solution
reads 
\begin{eqnarray}
(6\pi )^{-2}g_{2L}(t) &=&-\frac{1}{3t}+\frac{6a^{2}}{t^{5/3}}+\frac{1}{2t^{2}%
}(72a^{3}\ln t+27a^{3}+b)-\frac{1566a^{4}}{t^{7/3}}-\frac{3a^{2}}{t^{8/3}}%
(360a^{3}\ln t+2907a^{3}+5b)  \nonumber \\
&&\phantom{CCCCCCCCC}-3888a^{6}\frac{\ln ^{2}t}{t^{3}},  \nonumber \\
(6\pi )^{-2}\lambda (t) &=&\frac{a}{t^{4/3}}+\frac{9a^{2}}{t^{5/3}}+\frac{%
63a^{3}}{t^{2}}-\frac{a}{t^{7/3}}(144a^{3}\ln t+2b-495a^{3})  \nonumber \\
&&\phantom{CCCCCCCCC}-\frac{a^{2}}{t^{8/3}}\left( 1620a^{3}\ln t-\frac{138267%
}{14}a^{3}+\frac{45}{2}b\right) ,  \label{ggoo} \\
(6\pi )^{-2}u(t) &=&-\frac{2}{3t}+\frac{a}{t^{4/3}}+\frac{15a^{2}}{t^{5/3}}+%
\frac{1}{t^{2}}(72a^{3}\ln t+b)-\frac{a}{t^{7/3}}\left( 144a^{3}\ln t+\frac{{%
24669}}{7}a{^{3}+2}b\right)  \nonumber \\
&&\phantom{CCCCCCCCC}-\frac{3a^{2}}{t^{8/3}}\left( 900a^{3}\ln t+\frac{70587%
}{14}a^{3}+\frac{25}{2}b\right) -7776a^{6}\frac{\ln ^{2}t}{t^{3}},  \nonumber
\end{eqnarray}
up to $\mathcal{O}(t^{-3}\ln t)$, where $a$ and $b$ are the arbitrary
constants.

The cubic corrections to the beta functions start with terms $\sim 1/t^{3}$,
so they give extra contributions of the form $\sim c_{i}/t^{2}$ to the
solutions, where $c_{i}$ are uniquely determined functions of $a$ and $b$.
The corrections do not affect the terms proportional to $(\ln t)/t^{2}$.
Thus, the complete solution has the form 
\begin{eqnarray}
(6\pi )^{-2}g_{2L}(t) &=&-\frac{1}{3t}+\frac{6a^{2}}{t^{5/3}}+\frac{1}{2t^{2}%
}(72a^{3}\ln t+27a^{3}+b+\xi _{1})+\mathcal{O}(t^{-7/3}\ln t),  \nonumber \\
(6\pi )^{-2}\lambda (t) &=&\frac{a}{t^{4/3}}+\frac{9a^{2}}{t^{5/3}}+\frac{%
63a^{3}+\xi _{2}}{t^{2}}+\mathcal{O}(t^{-7/3}\ln t),  \label{asg} \\
(6\pi )^{-2}u(t) &=&-\frac{2}{3t}+\frac{a}{t^{4/3}}+\frac{15a^{2}}{t^{5/3}}+%
\frac{1}{t^{2}}(72a^{3}\ln t+b+\xi _{3})+\mathcal{O}(t^{-7/3}\ln t), 
\nonumber
\end{eqnarray}
where $\xi _{i}$, $i=1,2,3$, are calculable numbers, depending on the cubic
corrections to the beta functions. The other beta functions of (\ref{ba})
give 
\[
(6\pi )^{-2}g_{1L}=-\frac{3a^{2}}{5t^{5/3}}+\frac{\xi _{4}}{t^{2}}+\mathcal{O%
}(t^{-7/3}),\qquad (6\pi )^{-2}g_{R}=-\frac{18a^{2}}{5t^{5/3}}+\frac{\xi _{5}%
}{t^{2}}+\mathcal{O}(t^{-7/3}), 
\]
where $\xi _{4,5}$ are calculable numbers.

We have thus found a two dimensional domain of asymptotic freedom.

\paragraph{One-loop degeneracies}

In special cases, not frequent in physical problems, the one-loop
coefficients $c_{ijk}$ can have degeneracies that make the expansions of
some couplings start from powers $t^{-1/n}$ instead of $1/t$, similarly to
what happens in (\ref{b2}) when $\beta _{1}=0$ for a single coupling. Some
higher loop contributions can be as important as the one-loop ones, or even
more important than the one-loop ones. Then the expansions of the beta
functions in powers of the couplings have to be accordingly reordered. For
example, consider the system 
\[
\dot{g}_{1}=g_{1}^{2}+\kappa ^{2}g_{1}g_{2}^{2},\qquad \dot{g}_{2}=\frac{1}{4%
}g_{1}g_{2}, 
\]
where $\kappa $ is a constant. Observe that the terms $g_{1}^{2}$ and $%
\kappa ^{2}g_{1}g_{2}^{2}$ are one- and two-loop, respectively.
Nevertheless, they are equally important in the asymptotic expansion. The
Zimmermann trajectories are 
\begin{equation}
(g_{1},g_{2})=\left( -\frac{1}{t},0\right) ,\left( -\frac{2}{t},\pm \frac{1}{%
\kappa \sqrt{t}}\right) ,  \label{trage}
\end{equation}
to which we must add the line of fixed points $g_{1}\equiv 0$. The procedure
described above has to be applied using the trajectories (\ref{trage}). For
example, expanding around the second pair of trajectories we find that the
matrix $\mathcal{N}$ has eigenvalue 1 with degeneracy 2 and the expansions
read 
\begin{eqnarray*}
g_{1}(t) &=&-\frac{1}{t}\left( 2+\frac{a-b}{t}\ln t-\frac{a}{t}\right) +%
\mathcal{O}(t^{-3}\ln ^{2}t), \\
g_{2}(t) &=&\pm \frac{1}{4\kappa \sqrt{t}}\left( 4+\frac{a-b}{t}\ln t-\frac{b%
}{t}\right) +\mathcal{O}(t^{-5/2}\ln ^{2}t),
\end{eqnarray*}
where $a$ and $b$ are the arbitrary constants. All higher-order terms are
uniquely determined. The domain of asymptotic freedom is two-dimensional.

\paragraph{Another way to find $\mathcal{D}_{\text{AF}}$}

Here we give an alternative method that can be useful to determine the
domain $\mathcal{D}_{\text{AF}}$ of asymptotic freedom when the origin is an
isolated fixed point (possibly after a suitable projection in parameter
space). We define the radius $\rho $ in parameter space as 
\[
\rho =\sqrt{\sum_{i=1}^{N}g_{i}^{2}}, 
\]
and the radial velocity $v$ as 
\[
v=\frac{\mathrm{d}\rho }{\mathrm{d}t}=\frac{1}{\rho }\sum_{i=1}^{N}g_{i}%
\beta _{i}. 
\]
Let $\mathcal{D}$ denote a domain in parameter space and $\overline{\mathcal{%
D}}$ its closure. A theory is asymptotically free in $\mathcal{D}_{\text{AF}%
} $ if the origin belongs to $\mathcal{\bar{D}}_{\text{AF}}$ (but not $%
\mathcal{D}_{\text{AF}}$) and every trajectory passing through $\mathcal{D}_{%
\text{AF}}$ remains in $\mathcal{D}_{\text{AF}}$ and flows to the origin in
the ultraviolet limit. The trajectories that satisfy $v<0$ \textit{%
asymptotically }for $t\rightarrow \infty $ in a neighborhood of the origin
belong to $\mathcal{D}_{\text{AF}}$. The Zimmermann trajectories, in
particular, belong to $\mathcal{D}_{\text{AF}}$.

When the origin $g=0$ is an isolated fixed point $\mathcal{D}_{\text{AF}}$
can be also characterized as follows:

$i$) find the domain $\mathcal{D}$ around the origin where $v<0$;

$ii$) call $\partial _{0}\mathcal{D}$ the boundary of $\mathcal{D}$ minus
the origin, and consider the trajectories crossing it;

$iv$) if all such trajectories enter $\mathcal{D}$, then $\mathcal{D}_{\text{%
AF}}=\mathcal{D}$; if not, $\mathcal{D}_{\text{AF}}$ is $\mathcal{D}$ minus
the trajectories leaving $\mathcal{D}$ through the boundary $\partial _{0}%
\mathcal{D}$.

Indeed, the remaining trajectories cannot leave $\mathcal{D}$, so they must
flow to the origin, because it is the unique fixed point.

The condition $v<0$ is necessary, but not sufficient, because some
trajectories intersecting $\mathcal{D}$ can cross its boundary and run away,
instead of flowing to the origin. The good feature of $\mathcal{D}$ is that
it can be easily determined, but $\mathcal{D}_{\text{AF}}$ is only a subset
of $\mathcal{D}$.

Now, observe that $\mathcal{D}$ depends on the parametrization of the
couplings, while $\mathcal{D}_{\text{AF}}$ of course does not. Call $%
h_{i}(g) $ a reparametrization of the couplings and $\mathcal{D}_{h}$ the
domain where the velocity 
\[
v_{h}\equiv \frac{\sum_{i=1}^{N}h_{i}\beta _{h_{i}}}{\sqrt{%
\sum_{j=1}^{N}h_{j}^{2}}} 
\]
is negative in a neighborhood of the origin. An efficient way to estimate $%
\mathcal{D}_{\text{AF}}$ (and in most cases determine it) is to take the
intersection of the $\mathcal{D}_{h}$'s, for all reparametrizations $h$.

We illustrate this method in the $g_{2}$-$g_{4}$ subsystem of the Dirac
model in the large $N$ limit, using one-loop truncated beta functions.
Observe that the origin is the unique fixed point. We have 
\[
\rho v=g_{2}\beta _{2}+g_{4}\beta _{4}=ag_{2}(g_{2}^{2}+3g_{4}^{2}).
\]
The domain $\mathcal{D}$ is just $g_{2}<0$. Consider now the one-parameter
family of reparametrizations 
\[
h_{2}=g_{2}+\alpha g_{4},\qquad h_{4}=\alpha g_{2}+g_{4}.
\]
We find 
\[
\rho _{h}v_{h}=\frac{g_{2}^{3}}{r^{3}}\left[ r(1+\alpha
^{2})(3+r^{2})+2\alpha (1+3r^{2})\right] ,
\]
where $r=g_{2}/g_{4}$. Now we study the condition $v_{h}<0$, knowing that $%
g_{2}$ must be negative. Varying $\alpha $ to obtain the best result, we
find 
\[
|r|<1,\qquad \text{i.e.}\qquad |g_{2}|<|g_{4}|,
\]
which together with $g_{2}<0$ gives our best estimate of the domain $%
\mathcal{D}_{\text{AF}}$.

We can check this estimate using the explicit one-loop solution (\ref{11}).
We see that inside $\mathcal{D}$ only the trajectories with $%
|g_{2}|\leqslant |g_{4}|$ flow to the origin. All others cross the boundary $%
\partial _{0}\mathcal{D}$ (that is the line $g_{2}=0$), enter the region
with $v>0$ and run away. Thus, $\mathcal{D}_{\text{AF}}$ is given by $%
g_{2}<0 $, $|g_{2}|\leqslant |g_{4}|$. We conclude that our method gives an
accurate estimate of $\mathcal{D}_{\text{AF}}$, since it misses only its
boundary, namely the trajectories with $|g_{2}|=|g_{4}|$.

\section{Zimmermann trajectories and hidden symmetries}

\setcounter{equation}{0}

In the previous section we have seen that the Zimmermann trajectories play
an important role in the study of the domain of asymptotic freedom. Some
such trajectories (see for example (\ref{z2})) involve only rational
coefficients, others very complicated irrational numbers. Normally, rational
trajectories appear when the theory has more symmetries. In this section we
point out that the existence of rational Zimmermann trajectories appears to
be a general feature of high-energy Lorentz violating four fermion models.

First, we briefly recall Zimmermann's ``reductions of couplings'' \cite
{zimme,oheme2}. Assume that a theory has couplings $\lambda _{I}$, $%
I=1,\ldots N$. Zimmermann's idea is to parametrize the couplings in terms of
a smaller set of independent parameters $\alpha _{j}$, $j=1,\ldots M<N$.
Write $\lambda _{I}=\lambda _{I}(\alpha _{j})$. Consistence with the
renormalization group demands 
\[
\beta _{I}=\frac{\partial \lambda _{I}}{\partial \alpha _{j}}\beta _{j}. 
\]
The most interesting case is $N=2$, $M=1$. Normally, if the one-loop beta
functions are quadratic in the couplings, as in our case, Zimmermann's
equations admit two power-series solutions of the form 
\[
\bar{\lambda}=c\alpha +\alpha \sum_{n=1}^{\infty }d_{n}\alpha ^{n}, 
\]
where $c$ and $d_{k}$ are calculable, generically irrational, numbers. If
the solution exists at one loop (namely, if $c$ is real), then it exists to
all orders. The most general solution to Zimmermann's equations is not
analytic, but has the form 
\[
\bar{\lambda}^{\prime }=\bar{\lambda}+\sum_{m,n=1}^{\infty }d_{mn}\alpha
^{m\xi +n}, 
\]
where $\xi $ is typically irrational, $d_{11}$ is arbitrary and the other
coefficients $d_{mn}$ are uniquely determined once $d_{11}$ is given.

In special situations, such as when the ``reduced'' model has additional
symmetries, the power-series solution $\bar{\lambda}$ shrinks to the
monomial $c\alpha $, with a rational coefficient $c$, and $\bar{\lambda}%
^{\prime }$ becomes analytic.

For example, the (Lorentz invariant) model of a spinor $\psi $ and a
pseudoscalar field $A$ interacting with the Lagrangian 
\[
\mathcal{L}_{I}=igA\bar{\psi}\gamma _{5}\psi -\frac{\lambda }{4!}A^{4} 
\]
admits reductions 
\[
\lambda ^{\prime }=\frac{1}{3}(1\pm \sqrt{145})g^{2}+d_{1}g^{4}+\cdots
+d_{11}g^{\frac{2}{5}\sqrt{145}+2}+\cdots 
\]
On the other hand, the massless model with interaction 
\[
\mathcal{L}_{I}=g\bar{\psi}(A+i\gamma _{5}B)\psi -\frac{\lambda }{2}%
(A^{2}+B^{2})^{2} 
\]
admits the rational reduction 
\begin{equation}
\lambda =g^{2},  \label{lg2}
\end{equation}
which reveals the existence of a symmetry. Indeed, when the couplings are
related by formula (\ref{lg2}) we have the supersymmetric Wess-Zumino model.

We now analyze Zimmermann's trajectories at one loop in some of our Lorentz
violating four fermion models. Consider again the electroweak model. We look
for RG trajectories where all couplings are proportional to one another: $%
g_{1L}=ax$, $g_{2L}=bx$, $g_{R}=cx$, $g_{LR}=dx$ and $\lambda =ex$. The
constants $a$, $b$, $c$, $d$ and $e$ can be worked out matching the beta
functions (\ref{ba}). The absolute value of $b_{L}$ (or $b_{R}$) can be set
to 1 rescaling the space coordinates. On the other hand, the ratio $%
b_{L}/b_{R}$ is free and does not run at one loop.

We choose $|b_{L}|=|b_{R}|=1$ and consider the cases $s=\pm 1$. We find only
three real solutions. One of them is just $g_{1L}=g_{R}=g_{LR}=\lambda =0$,
with only $g_{2L}$ non-vanishing. Of the other two solutions, only one has
rational coefficients, and reads 
\begin{equation}
g_{1L}=\frac{\lambda }{5},\qquad g_{2L}=\lambda ,\qquad g_{R}=\frac{6}{5}%
\lambda ,\qquad g_{LR}=\frac{5+3s}{10}\lambda .  \label{myster}
\end{equation}
If we choose $|b_{L}|\neq |b_{R}|$ we generically do not find rational
trajectories.

The existence of a trajectory with rational coefficients is unexpected and
offers evidence that the reduced model might have hidden symmetries. Its
Lagrangian reads 
\begin{eqnarray*}
\mathcal{L}_{\mathrm{EWred}} &=&L^{\dagger i}i\left( \hat{\partial}+\mathbf{%
\sigma }\cdot \mathbf{\bar{\partial}}\frac{{\bar{\partial}}^{2}}{\Lambda
_{L}^{2}}\right) L^{i}+e_{R}^{\dagger }i\left( \hat{\partial}-s\mathbf{%
\sigma }\cdot \mathbf{\bar{\partial}}\frac{{\bar{\partial}}^{2}}{\Lambda
_{L}^{2}}\right) e_{R}-\frac{g}{\Lambda _{L}^{2}}(L^{\dagger
i}e_{R})(e_{R}^{\dagger }L^{i}) \\
&&-\frac{g}{10\Lambda _{L}^{2}}(L^{\dagger i}L^{i})^{2}-\frac{g}{2\Lambda
_{L}^{2}}(L^{\dagger i}L^{j})(L^{\dagger j}L^{i})-\frac{3g_{{}}}{5\Lambda
_{L}^{2}}(e_{R}^{\dagger }e_{R})^{2}-\frac{(5+3s)g}{10\Lambda _{L}^{2}}%
(L^{\dagger i}L^{i})(e_{R}^{\dagger }e_{R}),
\end{eqnarray*}
where $g=-\lambda >0$, and its one-loop beta function is 
\[
\beta _{g}=-\frac{5g^{2}}{(6\pi )^{2}}. 
\]

If the hidden symmetry is simple, we expect that the relations (\ref{myster}%
) are preserved by higher-loop corrections. However, this is not a necessary
requirement for a hidden symmetry.

The rational RG trajectory exists also in the $U(N_{L})\times U(N_{R})$
model for $|b_{L}|=|b_{R}|=1$. We find 
\[
g_{1L}=g_{1R}=-\frac{N_{L}+N_{R}-2}{2N_{L}+2N_{R}-1}g,\quad
g_{2L}=g_{2R}=g_{2LR}=-g,\quad g_{1LR}=-\frac{2N_{L}+2N_{R}+3s-1}{%
2(2N_{L}+2N_{R}-1)}g, 
\]
the beta function being 
\[
\beta _{g}=-\frac{g^{2}}{36\pi ^{2}}(2N_{L}+2N_{R}-1). 
\]

Curiously, the model of $N$ Dirac fermions in the large $N$ limit admits 
\textit{only} rational trajectories. We have already studied the $g_{2}$-$%
g_{4}$ subset in the previous section. In the $g_{5,6}$-$f_{1,2}$ subset we
find 15 rational trajectories: 
\begin{eqnarray*}
g_{5} &=&(-2,\pm 1,1,-1,-5,-2,-2,-7/5)g_{6}, \\
f_{1} &=&(0,0,\pm 1/2,\pm 1/2,\pm 3/2,\pm 1,\pm 1/2,\pm 1/10)g_{6}, \\
f_{2} &=&(0,0,0,\pm 1,\pm 3,\pm 3/2,\pm 3/2,0)g_{6}.
\end{eqnarray*}
One such trajectory ($g_{5}=g_{6}=2f_{1}$, $f_{2}=0$) gives the Lorentz
violating ``Gross-Neveu'' model \cite{gn}, whose interaction reads 
\[
\frac{\lambda }{2\Lambda _{L}^{2}}(\bar{\psi}^{i}\psi ^{i})^{2}=\frac{%
\lambda }{\Lambda _{L}^{2}}(\ell _{1}^{\dagger i}\ell _{1}^{j})(\ell
_{2}^{\dagger i}\ell _{2}^{j})+\frac{\lambda }{\Lambda _{L}^{2}}(\ell
_{1}^{\dagger i}\ell _{2}^{j})(\ell _{2}^{\dagger i}\ell _{1}^{j})+\frac{%
\lambda }{2\Lambda _{L}^{2}}\left[ (\ell _{1}^{Ti}\varepsilon \ell
_{2}^{i})(\ell _{1}^{Tj}\varepsilon \ell _{2}^{j})+(\ell _{1}^{\dagger
i}\varepsilon \ell _{2}^{*i})(\ell _{1}^{\dagger j}\varepsilon \ell
_{2}^{*j})\right] , 
\]
which is renormalizable by the same argument used in (\ref{zuto}). In some
sense, this is an example of ``hidden symmetry'' associated with the
rational trajectory.

\section{Conclusions}

\setcounter{equation}{0}

If Lorentz symmetry is violated at high energies, then the Standard Model
admits a ultraviolet completion that is renormalizable despite it contains
four fermion vertices at the fundamental level. In this scenario, four
fermion models play a key role, because all other interactions, being
super-renormalizable, disappear at energies much higher than the scale of
Lorentz violation. In this paper we have studied the one-loop
renormalization of CPT-invariant Lorentz violating four fermion models and
their RG flows.

We have first considered the most general case, working out formulas for the
beta functions, and then analyzed particular models in detail. We have
formulated a method to determine the domain of asymptotic freedom expanding
the running couplings around the free fixed point. We emphasize that if the
four fermion model (\ref{he}) is asymptotically free then the entire Lorentz
violating Standard Model is.

Moreover, we found that the RG flow admits a number of special ``rational''
trajectories that, in the spirit of Zimmermann's reduction of couplings,
might hide some new symmetries.

\end{document}